\documentclass[11pt]{article}
\usepackage[brazil]{babel}
\usepackage{url}
\usepackage{amssymb,amsxtra}
\usepackage{amsmath,color,graphicx}
\usepackage{lineno}
\usepackage{graphicx}
\RequirePackage{geometry}
\usepackage{indentfirst}
\usepackage[breakable]{tcolorbox}
\usepackage{lineno,hyperref}
\usepackage{alltt}
\usepackage{amsfonts}
\usepackage{amsmath,amssymb}
\usepackage{multirow}
\usepackage{eqparbox}
\usepackage{upquote} 
    \usepackage{eurosym} 
    \usepackage[mathletters]{ucs} 
    \usepackage{fancyvrb} 
    \usepackage{grffile} 
\definecolor{incolor}{HTML}{303F9F}
    \definecolor{outcolor}{HTML}{D84315}
    \definecolor{cellborder}{HTML}{CFCFCF}
\definecolor{cellbackground}{HTML}{F7F7F7}
\definecolor{darkgreen}{rgb}{0.00,0.50,0.00}
\definecolor{purple}{rgb}{0.67,0.13,1.00}
\definecolor{lgray}{rgb}{0.60,0.60,0.60}

\begin{document}

\def\chaptername{}
\def\contentsname{Sum\'{a}rio}
\def\listfigurename{Figuras}
\def\listtablename{Tabelas}
\def\abstractname{Resumo}
\def\appendixname{Ap\^{e}ndice}
\def\refname{\large Refer\^{e}ncias bibliogr\'{a}ficas}
\def\bibname{Bibliografia}
\def\indexname{\'{I}ndice remissivo}
\def\figurename{\small Fig.~}
\def\tablename{\small Tab.~}
\def\pagename{\small Pag.}
\def\seename{veja}
\def\alsoname{veja tamb\'em}
\def\na{-\kern-.4em\raise.8ex\hbox{{\tt \scriptsize a}}\ }
\def\pa{\slash \kern-.5em\raise.1ex\hbox{p}\ }
\def\ro{-\kern-.4em\raise.8ex\hbox{{\tt \scriptsize o}}\ }
\def\no{n$^{\underline{\rm o}}$}

\setcounter{tocdepth}{3}

\clearpage
\pagenumbering{arabic}

\thispagestyle{empty}
\parskip 8pt

\vspace*{0.2cm}
\begin{center}
{\huge \bf Sobre a covari\^{a}ncia da equa\c{c}\~{a}o de d'Alembert: os casos do som e da luz }\\
\ \\
{\huge \bf On the covariance of the d'Alembert equation: the cases of sound and light}\\

\vspace*{1.0cm}
{\Large \bf \it Francisco Caruso;\,$^{1}$ Vitor Oguri\,$^{2}$ }\\[2.em]

{{$^{1}$ Centro Brasileiro de Pesquisas F\'{\i}sicas, Coordena\c{c}\~{a}o de F\'{\i}sica de Altas Energias, 22290-180, Rio de Janeiro, RJ, Brasil.}}

{{$^{2}$ Universidade do Estado do Rio de Janeiro, Instituto de F\'{\i}sica Armando Dias Tavares, 20550-900, Rio de Janeiro, RJ, Brasil.}}
\end{center}

\noindent \textbf{Resumo}

A covari\^{a}ncia da equa\c{c}\~{a}o de d'Alembert para fen\^{o}menos ac\'{u}sticos, descritos por ondas mec\^{a}nicas em uma ou tr\^{e}s dimens\~{o}es espaciais, sob as transforma\c{c}\~{o}es de Galileu, \'{e} demonstrada sem a necessidade de abandonar a hip\'{o}tese de que o tempo \'{e} absoluto na Mec\^{a}nica Cl\'{a}ssica. Isto s\'{o} \'{e} verdade se e somente se a velocidade de fase do som depende da velocidade do observador. Por outro lado, mostra-se tamb\'{e}m que a mesma equa\c{c}\~{a}o de d'Alembert \'{e} covariante sob as transforma\c{c}\~{o}es de Lorentz se e somente se a velocidade de fase da luz n\~{a}o depender do observador.

\noindent \textbf{Palavras-chave:} equa\c{c}\~{a}o de d'Alembert; transforma\c{c}\~{o}es de Galileu; Transforma\c{c}\~{o}es de Lorentz; propaga\c{c}\~{a}o do som; propaga\c{c}\~{a}o da luz.

\vspace*{0.7cm}

\noindent \textbf{Abstract}

The covariance of the d'Alembert equation for acoustic phenomena, described by mechanical waves in one or three spatial dimensions, under Galilean transformations, is demonstrated without the need to abandon the hypothesis that time is absolute in Classical Mechanics. This is true only if and only if the phase velocity of sound depends on the velocity of the observer. On the other hand, it is also shown that the same d'Alembert equation is covariant under Lorentz transformations if and only if the phase velocity of light does not depend on the observer.


\vfill

\newpage

\section{Introdu\c{c}\~{a}o} \label{intro}

\paragraph*{}
Originalmente, o matem\'{a}tico franc\^{e}s Jean le Rond d'Alembert, derivou, em 1747, uma equa\-\c{c}\~{a}o diferencial parcial, da qual se obt\'{e}m a solu\c{c}\~{a}o geral para a propaga\c{c}\~{a}o de onda uni\-di\-mensional, considerando o problema de uma corda vibrante~\cite{Dalembert-a,Dalembert-b,Dalembert-c}. Com o tempo, ficou conhecida pelo seu sobrenome e passou a ter v\'{a}rias aplica\c{c}\~{o}es f\'{\i}sicas em Ac\'{u}stica. De um ponto de vista mais pr\'{a}tico, \'{e} eficazmente utilizada no levantamento s\'{\i}smico e na previs\~{a}o do comportamento do mar em um \textit{tsunami}, ou ainda em diagn\'{o}sticos m\'{e}dicos detalhados usando tecnologia de ultrassom. Do ponto de vista da F\'{\i}sica B\'{a}sica, \'{e} importante ter consci\^{e}ncia do fato de, sendo a Ac\'{u}stica um ramo da Mec\^{a}nica e sendo $v_s \ll c$, em que $v_s$ \'{e} velocidade de fase do som e $c$, a da luz, deve-se esperar que a equa\c{c}\~{a}o de d'Alembert, aplicada aos fen\^{o}menos ac\'{u}sticos, seja \textit{covariante} em rela\c{c}\~{a}o \`{a}s transforma\c{c}\~{o}es de Galileu.\footnote{\,As transforma\c{c}\~{o}es de Galileu s\~{a}o as equa\c{c}\~{o}es bem conhecidas que relacionam as coordenadas espa\c{c}o-temporais em dois sistemas associados a referenciais inerciais distintos, quando a velocidade relativa entre os referenciais s\~{a}o muito menores que a velocidade da luz no v\'{a}cuo.} Cabe lembrar que a covari\^{a}ncia de uma equa\c{c}\~{a}o n\~{a}o deriva, necessariamente, da invari\^{a}ncia de seus termos. Diz-se de uma equa\c{c}\~{a}o que ela \'{e} \textit{covariante} com rela\c{c}\~{a}o a uma dada transforma\c{c}\~{a}o quando as mudan\c{c}as de seus termos s\~{a}o tais que a forma da equa\c{c}\~{a}o \'{e} mantida, mesmo que seus termos n\~{a}o sejam \textit{invariantes}.

Em 2023, foi afirmado em um trabalho publicado que, no caso da Ac\'{u}stica, a equa\c{c}\~{a}o de d'Alembert \textit{n\~{a}o \'{e}} covariante pelas transforma\c{c}\~{o}es de Galileu, levando os autores a modificar essas transforma\c{c}\~{o}es~\cite{Valbona}. Entretanto, em 2024, veio \`{a} luz outro artigo mostrando que isso n\~{a}o \'{e} verdade~\cite{Caruso}. Foi esse embate que serviu de motiva\c{c}\~{a}o para a elabora\c{c}\~{a}o desse artigo did\'{a}tico, ampliando a discuss\~{a}o.

Por outro lado, o f\'{\i}sico escoc\^{e}s James Clerk Maxwell, tal qual todos os seus predecessores e contempor\^{a}neos que se de\-di\-caram aos estudos da eletricidade e do magnetismo, buscou, de algum modo, compatibilizar a descri\c{c}\~{a}o desses fen\^{o}menos com um modelo mec\^{a}nico, que dependia da aceita\c{c}\~{a}o de um meio (fluido, para alguns e s\'{o}lido para outros) imponder\'{a}vel, mas el\'{a}stico, permeando todo o espa\c{c}o: o \textit{\'{e}ter lumin\'{\i}fero}~\cite{Lima}. A hip\'{o}tese revolucion\'{a}ria de Maxwell foi considerar que, mesmo na aus\^{e}ncia de mat\'{e}ria ponder\'{a}vel (no v\'{a}cuo), haveria uma corrente de deslocamento, an\'{a}loga a uma corrente de condu\c{c}\~{a}o em um metal, a qual estaria presente se o campo el\'{e}trico fosse vari\'{a}vel no tempo, dando origem a um campo magn\'{e}tico e acarretando perturba\c{c}\~{o}es no \'{e}ter. Foi a introdu\c{c}\~{a}o da corrente de deslocamento que consolidou a simetria das leis do Eletromagnetismo (no v\'{a}cuo) e permitiu a Maxwell inferir que os campos eletromagn\'{e}ticos n\~{a}o eram t\~{a}o-somente artif\'{\i}cios matem\'{a}ticos para expressar os fen\^{o}menos eletromagn\'{e}ticos. Eles se propagavam, mesmo no espa\c{c}o ``vazio'' (no \'{e}ter), de um modo que a corrente de deslocamento provocada pela varia\c{c}\~{a}o do campo el\'{e}trico produzia um campo magn\'{e}tico tamb\'{e}m vari\'{a}vel, o qual, pela lei de Faraday, gerava tamb\'{e}m um campo el\'{e}trico vari\'{a}vel e, dessa forma, a perturba\c{c}\~{a}o era transmitida atrav\'{e}s do \'{e}ter~\cite{Oguri}. Foi assim que Maxwell chegou a uma das s\'{\i}nteses mais espetaculares da F\'{\i}sica, publicando, em 1865, seu conjunto de equa\c{c}\~{o}es~\cite{Maxwell}.

Apesar de terem sido estabelecidas no contexto da exist\^{e}ncia de um \'{e}ter, as equa\c{c}\~{o}es de Maxwell, \`{a} semelhan\c{c}a das leis de Newton, s\~{a}o ora consideradas leis fundamentais da F\'{\i}sica que transcendem suas g\^{e}neses. As grandezas e conceitos envolvidos passam a ter significados que s\'{o} s\~{a}o interpretados apropriadamente a partir das pr\'{o}prias leis. No caso do Eletromagnetismo, expresso pelas equa\c{c}\~{o}es de Maxwell, a defini\c{c}\~{a}o dos campos eletromagn\'{e}ticos n\~{a}o necessita mais dos modelos mec\^{a}nicos utilizados por Faraday e Maxwell.

A partir do conjunto das equa\c{c}\~{o}es de Maxwell, pode-se mostrar que tanto as componentes dos campos el\'{e}trico $\vec E$ e magn\'{e}tico $\vec B$, como os potenciais escalar $\phi$ e vetorial $\vec A$ satisfazem \`{a} equa\c{c}\~{a}o de d'Alembert. Maxwell havia compreendido, assim, que a luz era uma onda ele\-tro\-magn\'{e}tica, que se propagava no v\'{a}cuo com velocidade constante $c =1/\sqrt{\mu_{_0}\epsilon_{_0}}$. Entre 1886 e 1889, o f\'{\i}sico alem\~{a}o Heinrich Rudolf Hertz conduziu uma s\'{e}rie de experimentos com um circuito ressonante tentando provar que as a\c{c}\~{o}es eletromagn\'{e}ticas se propagam no ar com velocidade finita, com os quais acabou mostrando que os efeitos de propaga\c{c}\~{a}o de ondas que observava eram manifesta\c{c}\~{o}es das ondas eletromagn\'{e}ticas previstas por Maxwell.

Al\'{e}m desse fato, merece ser recordado que o f\'{\i}sico alem\~{a}o Woldemar Voigt, em 1887, em um estudo sobre o princ\'{\i}pio do efeito Doppler, demandou a covari\^{a}ncia da equa\c{c}\~{a}o de d'Alembert para dois sistemas de refer\^{e}ncia inerciais, admitindo a invari\^{a}ncia da velocidade da luz para os dois observadores inerciais. Obteve, assim, um conjunto de transforma\c{c}\~{o}es espa\c{c}o-temporais diferente das transforma\c{c}\~{o}es de Lorentz \cite{Heras,Engelhardt}. N\~{a}o se pretende aqui discutir quem teve a prioridade em propor uma generaliza\c{c}\~{a}o das transforma\c{c}\~{o}es de Galileu para o caso de velocidades pr\'{o}ximas \`{a} da luz. O intuito aqui \'{e} apenas ressaltar, mais uma vez, a relev\^{a}ncia da equa\c{c}\~{a}o de d'Alembert para a F\'{\i}sica B\'{a}sica. Lorentz chegou \`{a}s suas transforma\c{c}\~{o}es de uma maneira \textit{ad hoc}, mas baseado no conjunto completo das equa\c{c}\~{o}es de Maxwell e n\~{a}o em sua previs\~{a}o de uma onda eletromagn\'{e}tica. Quanto a Voigt, mais tarde ele se referiu com essas palavras a esse seu trabalho~\cite{Bucherer}:
\begin{quotation}
\noindent \baselineskip=10pt {\small
\textit{
Tratava-se das aplica\c{c}\~{o}es do princ\'{\i}pio Doppler, que ocorrem em partes especiais, embora n\~{a}o apenas com base na teoria eletromagn\'{e}tica, mas com base na teoria el\'{a}stica da luz. Contudo, j\'{a} ent\~{a}o foram encontradas algumas das mesmas consequ\^{e}ncias, que, mais tarde, foram obtidas a partir da teoria eletromagn\'{e}tica.}}
\end{quotation}

Entretanto, requerer a covari\^{a}ncia da equa\c{c}\~{a}o de d'Alembert, independente da motiva\c{c}\~{a}o, foi uma atitude cient\'{\i}fica de Voigt que veio para ficar.

Nesse ponto, o que confunde muitos estudantes \'{e} como uma mesma equa\c{c}\~{a}o pode ser covariante por dois grupos de transforma\c{c}\~{o}es diferentes? No caso do som, a equa\c{c}\~{a}o, como j\'{a} mencionado, deve ser covariante por Galileu, enquanto que, para a luz, covariante por Lorentz. Assim que deve ser e assim que \'{e}. O que \'{e} um resultado muito interessante \'{e} que, para isso ocorrer, v\'{\i}nculos s\~{a}o impostos sobre se a velocidade de fase da onda depende ou n\~{a}o da velocidade de um observador (inercial). \'{E} isso que ser\'{a} abordado nas pr\'{o}ximas Se\c{c}\~{o}es.

\section{A covari\^{a}ncia da equa\c{c}\~{a}o de d'Alembert para as ondas sonoras: o caso unidimensional} \label{acustica-1D}

Em um primeiro momento, mostra-se a covari\^{a}ncia da equa\c{c}\~{a}o de d'Alembert para fen\^{o}\-me\-nos ac\'{u}sticos, sob as transforma\c{c}\~{o}es de Galileu, em uma dimens\~{a}o espacial. A prova para tr\^{e}s dimens\~{o}es \'{e} apresentada em sequ\^{e}ncia (Se\c{c}\~{a}o~\ref{acustica-3D}).

Com rela\c{c}\~{a}o a um sistema cartesiano de coordenadas $S$, associado a um referencial estacion\'{a}rio em um meio n\~{a}o dispersivo, a equa\c{c}\~{a}o de onda que rege a propaga\c{c}\~{a}o de uma onda sonora ao longo de um de seus eixos de coordenadas, por exemplo, o eixo $x$, \'{e} dada por
$$ \displaystyle \frac{1}{v_s^2} \, \frac{\partial^2}{\partial t^2} \Psi(x,t) = \frac{\partial^2}{\partial x^2} \Psi(x,t) $$

\noindent
em que $v_s$ \'{e} a velocidade de fase do som,\footnote{\, Para um referencial estacion\'{a}rio no meio de propaga\c{c}\~{a}o, velocidade de fase do som n\~{a}o depende do movimento da fonte, a fonte estabelece apenas a frequ\^{e}ncia da onda gerada. Toda onda harm\^{o}nica se propaga com a velocidade de fase constante $v_s$, em rela\c{c}\~{a}o ao meio.} $(x, t)$ s\~{a}o as coordenadas espacial e temporal de um ponto do meio, segundo $S$, e $\Psi(x,t)$ \'{e} um campo escalar,\footnote{\, Genericamente, denominado fun\c{c}\~{a}o de onda, como todo campo que obedece \`{a} equa\c{c}\~{a}o de onda.} como a press\~{a}o ou a densidade do meio, cujo valor, em um dado instante e ponto do meio, n\~{a}o depende do referencial.

Assim, se $(x^\prime, t^\prime)$ s\~{a}o as coordenadas espa\c{c}o-temporais de um ponto do meio, com rela\c{c}\~{a}o a um sistema $S^\prime$ associado a um referencial n\~{a}o estacion\'{a}rio, com eixos paralelos a $S$, que se desloca no sentido positivo do eixo $x$ de $S$, tal que no instante inicial $t^\prime = t =0$, a equa\c{c}\~{a}o seguinte deve ser v\'{a}lida.
\begin{equation} \label{invar_psi}
\Psi (x,t) = \Psi^\prime (x^\prime, t^\prime)
\end{equation}

Sabe-se tamb\'{e}m que, a forma geral da fun\c{c}\~{a}o de onda, $\Psi (x,t)$, de um pulso que se propaga em um meio n\~{a}o dispersivo, no sentido positivo do eixo $x$ de $S$, \'{e} dada por
$$ \Psi (x,t) = f (x - v_s t) $$
Al\'{e}m disso, as transforma\c{c}\~{o}es de Galileu, entre as coordenadas em $S$ e $S^\prime$,
$$ x = x^\prime + V t^\prime$$
pressup\~{o}em um tempo absoluto e implicam as seguintes rela\c{c}\~{o}es entre as derivadas espaciais e temporais,
$$ \left\{
\begin{array}{l}
\displaystyle
\frac{\partial}{\partial x^\prime} = \overbrace{\frac{\partial x}{\partial x^\prime}}^{1} \frac{\partial}{\partial x}
\quad \Rightarrow \quad \frac{\partial^2}{\partial {x^\prime}^2} = \frac{\partial^2}{\partial x^2} \\
 \ \\
\displaystyle
\frac{\partial}{\partial t^\prime} = \overbrace{\frac{\partial t}{\partial t^\prime}}^{1} \frac{\partial}{\partial t} \ + \ \overbrace{\frac{\partial x}{\partial t^\prime}}^{V}  \frac{\partial}{\partial x} \quad \Rightarrow \quad
 \frac{\partial^2}{\partial {t^\prime}^2} = \frac{\partial^2}{\partial t^2} + 2 V \frac{\partial^2}{\partial t \partial x} + V^2 \frac{\partial^2}{\partial x^2}
  \end{array}
\right.
$$
Desse modo, tem-se
$$ \left\{
\begin{array}{l}
\displaystyle
\frac{\partial^2}{\partial {x^\prime}^2} \Psi^\prime(x^\prime, t^\prime) = \frac{\partial^2}{\partial x^2} \Psi(x, t)  \\
 \ \\
\displaystyle
\frac{\partial^2}{\partial {t^\prime}^2} \Psi^\prime(x^\prime, t^\prime) = \frac{\partial^2}{\partial t^2} \Psi(x, t) + 2 V \frac{\partial^2}{\partial xt \partial x} \Psi(x, t) + V^2 \frac{\partial^2}{\partial x^2} \Psi(x, t)
  \end{array}
\right.
$$

Levando-se em conta a forma geral da fun\c{c}\~{a}o de onda, obt\'{e}m-se:
$$ \Psi(x,t) = f \underbrace{(x - v_s t)}_{y} \quad \Rightarrow \quad
\left\{
\begin{array}{l}
\displaystyle
\frac{\partial^2}{\partial t^2} \Psi(x, t) = v_s^2 f^{\prime \prime} \\
\ \\
\displaystyle
\frac{\partial^2}{\partial t \partial x} \Psi(x, t) = - v_s f^{\prime \prime} \\
\ \\
\displaystyle
\frac{\partial^2}{\partial x^2} \Psi (x,t) = f^{\prime \prime}
\end{array}
\right.
$$
na qual $f^{\prime \prime} = \mbox{d}^2f/\mbox{d}y^2$. Com esse resultado, a equa\c{c}\~{a}o de d'Alembert se escreve
$$
\displaystyle
\frac{\partial^2}{\partial {t^\prime}^2} \Psi^\prime(x^\prime, t^\prime) = \big( \underbrace{v_s^2 - 2 v_s V + V^2}_{(v_s - V)^2} \big) f^{\prime \prime} = {v_s^\prime}^2 \, \frac{\partial^2}{\partial {x^\prime}^2} \Psi^\prime(x^\prime, t^\prime)
$$
ou seja,
$$
\fbox{~~~
\begin{minipage}{5.8cm}
\vspace*{0.15cm}
$\displaystyle \frac{1}{{v_s^\prime}^2} \, \frac{\partial^2}{\partial {t^\prime}^2} \Psi^\prime(x^\prime, t^\prime) = \frac{\partial^2}{\partial {x^\prime}^2} \Psi^\prime(x^\prime, t^\prime)$
\vspace*{0.15cm}
\end{minipage}
~}
$$
sendo $v_s^\prime = v_s - V$, a velocidade de fase do som, segundo $S^\prime$.

Assim, se a velocidade de fase do som depende da velocidade do referencial, a equa\c{c}\~{a}o de ondas espacialmente unidimensional tem a mesma forma em qualquer referencial inercial.



Esse resultado mostra que, no caso unidimensional,  a velocidade de fase do som, segundo um referencial n\~{a}o estacion\'{a}rio no meio, \'{e} alterada de modo similar \`{a} composi\c{c}\~{a}o de velocidades de part\'{\i}culas. No entanto, \'{e} oportuno enfatizar que a velocidade de fase  n\~{a}o \'{e} a velocidade de qualquer objeto ponder\'{a}vel e, portanto, n\~{a}o deve ser tratada como uma grandeza cinem\'{a}tica.

Esse ponto pode ser esclarecido, ao se considerar a propaga\c{c}\~{a}o de um onda sonora em uma dire\c{c}\~{a}o diferente de qualquer um dos eixos de um sistema cartesiano de coordenadas.

\section{A covari\^{a}ncia da equa\c{c}\~{a}o de d'Alembert para as ondas sonoras: o caso tridimensional} \label{acustica-3D}

Com rela\c{c}\~{a}o a um sistema cartesiano de coordenadas $S$, associado a um referencial  estacion\'{a}rio em um meio n\~{a}o dispersivo, a propaga\c{c}\~{a}o de ondas sonoras no espa\c{c}o obedece \`{a} equa\c{c}\~{a}o (tridimensional) de d'Alembert, dada por
$$ \displaystyle \left( \frac{\partial^2}{\partial x^2} + \frac{\partial^2}{\partial y^2} + \frac{\partial^2}{\partial z^2} \right) \Psi(x,y,z,t) = \nabla^2\, \Psi(x,y,z,t) = \frac{1}{v_s^2} \, \frac{\partial^2}{\partial t^2} \Psi(x,y,z,t) $$
em que ($x,y,z,t$) s\~{a}o as coordenadas espaciais e temporal, $\nabla^2 = \frac{\partial^2}{\partial x^2} + \frac{\partial^2}{\partial y^2}+ \frac{\partial^2}{\partial z^2}$ \'{e} o operador laplaciano, $\Psi(x,y,z,t)$ representa a press\~{a}o ou a densidade do meio, e $v_s$ \'{e} a velocidade de fase do som.

\begin{figure}[htbp]
\centerline{\includegraphics[width=5.5cm]{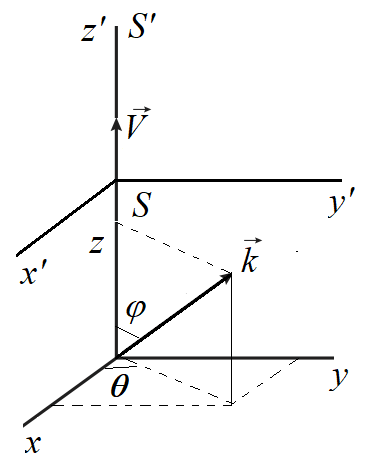}}
\caption{\baselineskip=6pt
{\small \textbf{Sistemas de coordenadas $S$ e $S^\prime$. $V$ \'{e} a velocidade de $S^\prime$. }}}
\label{onda_tri}
\end{figure}

Assim, se $(x^\prime, y^\prime, z^\prime, t^\prime)$ s\~{a}o as coordenadas espa\c{c}o-temporais com rela\c{c}\~{a}o a um sistema $S^\prime$ associado a um referencial n\~{a}o estacion\'{a}rio (Figura~\ref{onda_tri}), mas com eixos paralelos a $S$, que se desloca no sentido positivo do eixo $z$ de $S$, com velocidade $V$, tal que no instante inicial $t^\prime = t= 0$, a equa\c{c}\~{a}o seguinte deve ser v\'{a}lida.
\begin{equation} \label{invar_psi}
\Psi (x,y,z,t) = \Psi^\prime (x^\prime, y^\prime, z^\prime, t^\prime)
\end{equation}

Em rela\c{c}\~{a}o a $S$, uma onda harm\^{o}nica que se propaga com o vetor de propaga\c{c}\~{a}o $\vec k$, indicado na Figura~\ref{onda_tri}, pode ser expressa como
$$ \Psi (x,y,z,t) = A \, \mbox{sen}\, \big( k_x x + k_y y + k_z z - \omega t \big) $$
sendo $k_x = k \, \mbox{sen}\, \varphi \cos \theta$, $k_y = k \, \mbox{sen}\, \varphi \, \mbox{sen}\, \theta $ e $k_z = k \cos \varphi $, as componentes de $\vec k$, cuja  magnitude \'{e} igual a $ k = \omega/v_s$.
Uma vez que, no dom\'{\i}nio n\~{a}o relativ\'{\i}stico, as coordenadas em $S$ e $S^\prime$ est\~{a}o relacionadas pelas trans\-for\-ma\c{c}\~{o}es de Galileu, como
$$ x = x^\prime, \qquad y = y^\prime, \qquad z = z^\prime + V t^\prime \qquad \mbox{e} \qquad t = t^\prime$$
%
\noindent as rela\c{c}\~{o}es entre as derivadas espaciais e temporais nos sistemas $S$ e $S^\prime$,
$$ \displaystyle \frac{\partial^2}{\partial {x^\prime}^2} = \frac{\partial^2}{\partial x^2}, \qquad \frac{\partial^2}{\partial {y^\prime}^2} = \frac{\partial^2}{\partial y^2}, \qquad \frac{\partial^2}{\partial {z^\prime}^2} = \frac{\partial^2}{\partial z^2} \qquad \mbox{e} \qquad \frac{\partial^2}{\partial {t^\prime}^2} = \frac{\partial^2}{\partial t^2} + 2 V \frac{\partial^2}{\partial t \partial z} + V^2 \frac{\partial^2}{\partial z^2} $$
mostram que $ \nabla^2 = {\nabla^\prime}^2$.

Expressando-se a fun\c{c}\~{a}o de onda no sistema $S$ em termos dos  \^{a}ngulos $\theta$ e $\varphi$,
{ $$ \Psi (x,y,z,t) = A \, \mbox{sen}\, k \Big[ \big( \underbrace{\mbox{sen}\, \varphi \cos \theta}_{\alpha} x + \, \underbrace{\mbox{sen}\, \varphi \, \mbox{sen}\, \theta}_{\beta} y + \underbrace{\cos \varphi}_{\gamma} z \big) - v_s t \Big] = f (\alpha x + \beta y + \gamma z - v_s t)$$ }
e tendo em conta que
$$
\displaystyle
\frac{\partial^2}{\partial x^2} \Psi = \alpha^2 f^{\prime \prime}, \qquad
\frac{\partial^2}{\partial y^2} \Psi = \beta^2  f^{\prime \prime}, \qquad
\frac{\partial^2}{\partial z^2} \Psi = \gamma^2 f^{\prime \prime} \qquad \mbox{e} \qquad
\frac{\partial^2}{\partial t \partial z} \Psi = - \gamma v_s f^{\prime \prime}
$$
resulta
$$ \nabla^2 \Psi = {\nabla^\prime}^2 \Psi^\prime = \big( \underbrace{\alpha^2 + \beta^2 + \gamma^2}_{1} \big) f^{\prime \prime} $$

Assim, obt\'{e}m-se
$$ \displaystyle \frac{\partial^2}{\partial {t^\prime}^2} \Psi^\prime = \big( v_s^2 - 2 \gamma v_s V + \gamma^2 \big) f^{\prime \prime} = \big( \underbrace{ v_s^2 - 2 v_s V \cos \varphi + V^2 \cos^2 \varphi}_{(v_s - V \cos \phi)^2} \big) {\nabla^\prime}^2 \Psi^\prime$$
ou seja,
$$
\fbox{~~~
\begin{minipage}{7.3cm}
\vspace*{0.15cm}
$\displaystyle \displaystyle {\nabla^\prime}^2 \Psi^\prime (x^\prime,y^\prime,z^\prime,t^\prime) = \frac{1}{{v_s^\prime}^2}
\frac{\partial^2}{\partial {t^\prime}^2} \Psi^\prime (x^\prime,y^\prime,z^\prime,t^\prime) $
\vspace*{0.15cm}
\end{minipage}
~}
$$
sendo $v_s^\prime = (v_s - V \cos \varphi)$ a velocidade de fase do som, segundo $S^\prime$.

Desse modo, pode-se concluir que a equa\c{c}\~{a}o de propaga\c{c}\~{a}o das ondas sonoras tem a mesma forma em qual\-quer referencial inercial, se e somente se a velocidade de fase do som depender da velocidade do referencial. Alternativamente, essa propriedade \'{e} expressa dizendo-se que a equa\c{c}\~{a}o de d'Alembert para ondas ac\'{u}sticas \'{e} \textit{covariante} com rela\c{c}\~{a}o \`{a}s transforma\c{c}\~{o}es de Galileu.

Esse resultado, no caso tridimensional, mostra sem d\'{u}vidas que a velocidade de fase do som, segundo um referencial n\~{a}o estacion\'{a}rio no meio, n\~{a}o \'{e} alterada de modo similar \`{a} composi\c{c}\~{a}o de velocidades de part\'{\i}culas.\footnote{\, Em geral, quando a propaga\c{c}\~{a}o da onda n\~{a}o tem a mesma dire\c{c}\~{a}o que a velocidade de transla\c{c}\~{a}o do referencial n\~{a}o estacion\'{a}rio no meio ($S^\prime$), essa similaridade n\~{a}o ocorre.} Ou seja, a velocidade de fase do som n\~{a}o obedece \`{a} regra de composi\c{c}\~{a}o de velocidades da cinem\'{a}tica newtoniana. Assim, a abordagem tridimensional, al\'{e}m da covari\^{a}ncia da equa\c{c}\~{a}o de onda sonora, mostra que de fato a velocidade de fase n\~{a}o \'{e} uma grandeza vetorial.

\section{A covari\^{a}ncia da equa\c{c}\~{a}o de d'Alembert para as ondas luminosas}\label{invar_luz}

Sejam dois referenciais inerciais ($S$ e $S^\prime$), tal que
$$
\left\{
\begin{array}{l}
x^\prime = \gamma(V)\ (x - Vt) \\
\ \hspace{5cm} \gamma = \big( 1 - V^2/c^2 \big)^{-1/2} \\
t^\prime = \gamma(V)\ (t - Vx/c^2)
\end{array}
\right.
$$
em que $c$ \'{e} a velocidade de fase da luz para o referencial $S$ e $V$, a velocidade relativa entre eles.

Para a luz, com base nas equa\c{c}\~{o}es de Maxwell, pode-se mostrar que as componentes dos campos $\vec E$ e $\vec B$, bem como os potenciais escalar $\phi$ e vetorial $\vec A$, satisfazem tamb\'{e}m a equa\c{c}\~{a}o de d'Alembert.
\index{equa\c{c}\~{a}o! de onda}
\index{equa\c{c}\~{a}o! de onda! de d'Alembert}
\index{d'Alembert! equa\c{c}\~{a}o de onda de}

\begin{figure}[htbp]
\centerline{\includegraphics[width=6.cm]{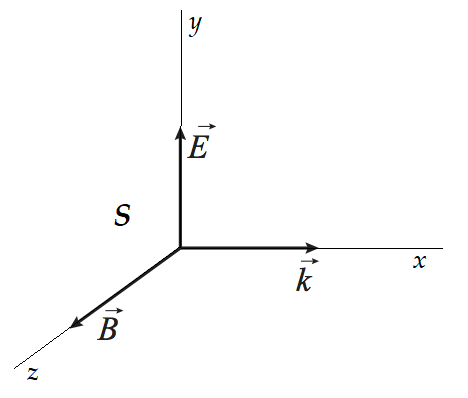}}
\caption{\baselineskip=10pt
{\small \textbf{Onda plana eletromagn\'{e}tica segundo $S$.}}}
\label{triedro}
\end{figure}

Seja uma onda eletromagn\'{e}tica plana que se propaga no sentido positivo do eixo $x$ (Figura.~\ref{triedro}) de um sistema de refer\^{e}ncia $S$, cujos campos $\vec E$ e $\vec B$ s\~{a}o dados por
\begin{equation}\label{campos_ondas}
\left\{
\begin{array}{l} \displaystyle
\vec E = \hat y \, \underbrace{E_{_0} e^{\displaystyle i ( k x - \omega t)}}_{\displaystyle E_y (x,t)} \\
\ \\
\displaystyle
\vec B = \hat z \, \underbrace{B_{_0} e^{\displaystyle i ( k x - \omega t)}}_{\displaystyle B_z (x,t)}
\end{array}
\right.
\end{equation}
e, portanto, suas componentes obedecem \`{a} equa\c{c}\~{a}o de onda em uma dimens\~{a}o,
\begin{equation}\label{dalemb-para-luz}
\displaystyle \left( \frac{\partial^2}{\partial x^2} \, - \, \frac{1}{c^2} \frac{\partial^2}{\partial t^2} \right)
\left( \begin{array}{c}
          E_y  \\
          B_z
        \end{array}
\right) = 0
\end{equation}

Esta equa\c{c}\~{a}o tem um papel decisivo na pr\'{o}pria interpreta\c{c}\~{a}o da Teoria de Maxwell, pois \'{e} ela que evidencia a natureza ondulat\'{o}ria da propaga\c{c}\~{a}o das ondas eletromagn\'{e}ticas e, em especial, da luz.

Admita-se, por ora, do mesmo modo como era esperado na \'{e}poca da realiza\c{c}\~{a}o do experimento de Michelson-Morley, que a velocidade de fase da luz, segundo o observador em $S^\prime$, que se desloca com velocidade constante em rela\c{c}\~{a}o ao referencial $S$, tenha um valor diferente $c^{\,\prime}$, por causa do vento do \'{e}ter~\cite{Oguri}. Logo, espera-se que nesse re\-ferencial,
\begin{equation}\label{dalemb-para-luz-linha}
\displaystyle \left( \frac{\partial^2}{\partial x^{\prime 2}}\, -\, \frac{1}{c^{\,\prime 2}} \frac{\partial^2}{\partial t^{\prime 2}} \right)
\left( \begin{array}{c}
          E_y^\prime \\
          B_z^\prime
        \end{array}
\right) = 0
\end{equation}
em que os campos dependem agora de $(x^\prime, t^\prime)$.

Relacionando-se as derivadas nos referenciais $S$ e $S^\prime$, ou seja
\begin{equation}\label{campos_derivadas}
\left\{
\begin{array}{l} \displaystyle
\displaystyle \frac{\partial}{\partial x} = \gamma \left(\frac{\partial}{\partial x^\prime} - \frac{V}{c^2}\frac{\partial}{\partial t^\prime}\right) \quad \Rightarrow \quad \frac{\partial^2}{\partial x^2} = \gamma^2 \left[ \frac{\partial^2}{\partial x^{\,\prime 2}} - 2 \frac{V}{c^2}\frac{\partial^2}{\partial x^\prime \partial t^\prime} + \frac{V^2}{c^4}\frac{\partial^2}{\partial t^{\,\prime 2}} \right] \\
  \ \\
  \displaystyle \frac{\partial}{\partial t} = \gamma \left(\frac{\partial}{\partial t^\prime} - V \frac{\partial}{\partial x^\prime}\right) \quad \Rightarrow \quad \frac{1}{c^2} \frac{\partial^2}{\partial t^2} = \gamma^2 \left[ \frac{1}{c^2} \frac{\partial^2}{\partial t^{\,\prime 2}} - 2\frac{V}{c^2} \frac{\partial^2}{\partial x^\prime \partial t^\prime} + \frac{V^2}{c^2} \frac{\partial^2}{\partial x^{\,\prime 2}} \right]
\end{array}
\right.
\end{equation}

Apesar de os campos eletromagn\'{e}ticos observados em $S$ n\~{a}o serem os mesmos do que aqueles observados em $S^\prime$, eles dependem linearmente uns dos outros, ou seja,
\begin{equation}\label{campos_trans}
\left\{
\begin{array}{l} \displaystyle
E_y (x, t) = a E_y^\prime (x^\prime, t^\prime) + b B_z^\prime (x^\prime, t^\prime) \\
 \ \\
 \displaystyle
 B_z (x, t) = d E_y^\prime (x^\prime, t^\prime) + e B_z^\prime (x^\prime, t^\prime)
\end{array}
\right.
\end{equation}

Substituindo as equa\c{c}\~{o}es~(\ref{campos_derivadas}) e (\ref{campos_trans}) na equa\c{c}\~{a}o~(\ref{dalemb-para-luz}), resulta

\begin{eqnarray*}
\displaystyle \left[ \frac{\partial^2}{\partial x^{\,\prime 2}} - 2 \frac{V}{c^2}\frac{\partial^2}{\partial x^\prime \partial t^\prime} + \frac{V^2}{c^4}\frac{\partial^2}{\partial t^{\,\prime 2}} \right] \big( a E_y^\prime + b B_z^\prime \big) &+& \, \\
- \left[ \frac{1}{c^2} \frac{\partial^2}{\partial t^{\,\prime 2}} - 2\frac{V}{c^2} \frac{\partial^2}{\partial x^\prime \partial t^\prime} + \frac{V^2}{c^2} \frac{\partial^2}{\partial x^{\,\prime 2}} \right] \big( a E_y^\prime + b B_z^\prime \big) = 0 &&
\end{eqnarray*}
ou,
$$  \displaystyle
a \left( 1 - \frac{V^2}{c^2} \right) \frac{\partial^2 E_y^\prime}{\partial x^{\,\prime 2}} +
b \left( 1 - \frac{V^2}{c^2} \right) \frac{\partial^2 B_z^\prime}{\partial x^{\,\prime 2}} -
\frac{a}{c^2} \left( 1 - \frac{V^2}{c^2} \right) \frac{\partial^2 E_y^\prime}{\partial t^{\,\prime 2}} -
\frac{b}{c^2} \left( 1 - \frac{V^2}{c^2} \right) \frac{\partial^2 B_z^\prime}{\partial t^{\,\prime 2}} = 0
$$

Assim, obt\'{e}m-se
$$ \displaystyle
a \left( \frac{\partial^2 }{\partial x^{\,\prime 2}} - \frac{1}{c^2} \frac{\partial^2 }{\partial t^{\,\prime 2}} \right) E_y^\prime +
b \left( \frac{\partial^2 }{\partial x^{\,\prime 2}} - \frac{1}{c^2} \frac{\partial^2 }{\partial t^{\,\prime 2}} \right) B_z^\prime = 0 \quad \Longrightarrow \quad
\displaystyle \left( \frac{\partial^2}{\partial x^{\prime 2}}\, -\, \frac{1}{c^2} \frac{\partial^2}{\partial t^{\prime 2}} \right)
\left( \begin{array}{c}
          E_y^\prime  \\
          B_z^\prime
        \end{array}
\right) = 0
$$

Dessa maneira, em rela\c{c}\~{a}o ao sistema $S^\prime$, as equa\c{c}\~{o}es de propaga\c{c}\~{a}o para as componentes dos campos ele\-tro\-magn\'{e}ticos  s\'{o} mant\^{e}m a mesma forma se e somente se $c^\prime = c$. Ou seja, se a velocidade de fase de uma onda eletromagn\'{e}tica no v\'{a}cuo \'{e} uma constante $c$, in\-de\-pen\-den\-te\-mente da fonte e do observador.

Portanto, a covari\^{a}ncia da equa\c{c}\~{a}o de onda de d'Alem\-bert para os fen\^{o}menos luminosos exige que a velocidade de fase da luz seja um invariante de Lorentz, princ\'{\i}pio adotado por Einstein, em 1905. Baseando-se nesse resultado, pode-se pensar que $c^\prime = c$  n\~{a}o seja um princ\'{\i}pio que ``estranhamente d\'{a} certo'', mas como uma condi\c{c}\~{a}o necess\'{a}ria para que, {n\~{a}o apenas a equa\c{c}\~{a}o de onda para os campos eletromagn\'{e}ticos, mas tamb\'{e}m as pr\'{o}prias equa\c{c}\~{o}es de Maxwell do Eletromagnetismo,\footnote{\, Das quais se deriva a equa\c{c}\~{a}o de onda.} e todas as leis e equa\c{c}\~{o}es fundamentais da F\'{\i}sica sejam covariantes nas mudan\c{c}as de referenciais inerciais, segundo as transforma\c{c}\~{o}es de Lorentz.

Deve-se, a este ponto, mencionar um detalhe epistemol\'{o}gico relevante, que pode nos levar a fazer uma leitura do resultado aqui obtido para a luz com certa cautela. Isso adv\'{e}m do fato de as transforma\c{c}\~{o}es de Lorentz terem sido constru\'{\i}das tendo o Eletromagnetismo de Maxwell como paradigma de teoria f\'{\i}sica. Naturalmente, pode-se questionar se, de alguma forma, ela n\~{a}o contempla a invari\^{a}ncia da velocidade da luz, como proposto no artigo~\cite{Pars}. Uma vez convencido dessa possibilidade, o resultado da Se\c{c}\~{a}o~\ref{invar_luz} deve ser entendido mais como uma verifica\c{c}\~{a}o de consist\^{e}ncia do que como uma prova. Espera-se discutir esse ponto em outra publica\c{c}\~{a}o.

Por outro lado, para o som, n\~{a}o cabe essa ressalva. As transforma\c{c}\~{o}es de Galileu n\~{a}o foram originalmente proposta tendo por base uma equa\c{c}\~{a}o de onda e sim a Mec\^{a}nica das part\'{\i}culas. A condi\c{c}\~{a}o imposta sobre a velocidade de fase da onda descrita pela equa\c{c}\~{a}o de d'Alembert dependa da velocidade do referencial \'{e} fundamental para que esta equa\c{c}\~{a}o possa ser tamb\'{e}m covariante pelas transforma\c{c}\~{o}es de Galileu.

\section{Coment\'{a}rios finais}

Que a velocidade de fase de propaga\c{c}\~{a}o de uma onda em um meio n\~{a}o dispersivo depende da velocidade do meio e do observador, enquanto a da luz, n\~{a}o, s\~{a}o fatos emp\'{\i}ricos. O que se mostrou neste artigo \'{e} que, do ponto de vista te\'{o}rico, \'{e} o conjunto de transforma\c{c}\~{o}es que deixam uma determinada equa\c{c}\~{a}o de onda covariante que intrinsecamente determina esse ou aquele comportamento ondulat\'{o}rio. Nos casos aqui considerados, tanto as transforma\c{c}\~{o}es de Galileu quanto de Lorentz s\~{a}o lineares e formam grupos. O que as difere \'{e} a estrutura b\'{a}sica do espa\c{c}o-tempo, com reflexos nas rela\c{c}\~{o}es entre as coordenadas para dois referenciais inerciais observando o mesmo fen\^{o}meno. No caso cl\'{a}ssico, admite-se um espa\c{c}o euclideano e um tempo absolutos, \textit{\`{a} la} Newton; no caso relativ\'{\i}stico, Einstein introduziu a relatividade entre o espa\c{c}o e tempo, que passam a n\~{a}o ser independentes, integrando, ent\~{a}o, um espa\c{c}o-tempo quadridimensional. Ao implementar isso, s\~{a}o revistos conceitos basilares como os de simultaneidade e causalidade e esses novos conceitos passam a estar, de alguma forma, incorporados nas leis de transforma\c{c}\~{a}o das coordenadas~\cite{Einstein}. Costuma-se dizer que a Relatividade, embora tendo sido elaborada tendo a covari\^{a}ncia do Eletromagnetismo como paradigma, acabou transformando-se em uma teoria para o espa\c{c}o-tempo. A esse respeito, veja o coment\'{a}rio do f\'{\i}sico estadunidense Arhur I. Miller~\cite{Miller}, citando o pr\'{o}prio Einstein,
\begin{quotation}
\noindent \baselineskip=10pt {\small
\textit{
A nova caracter\'{\i}stica [da teoria da relatividade de 1905] foi a compreens\~{a}o do facto de que a influ\^{e}ncia da transforma\c{c}\~{a}o de Lorentz transcendia a sua liga\c{c}\~{a}o com as equa\c{c}\~{o}es de Maxwell e estava preocupada com a natureza do espa\c{c}o e do tempo em geral. Um outro resultado novo foi que a ``invari\^{a}ncia de Lorentz'' \'{e} uma condi\c{c}\~{a}o geral para qualquer teoria f\'{\i}sica. Isto foi para mim de particular import\^{a}ncia porque eu j\'{a} tinha descoberto anteriormente que a teoria de Maxwell n\~{a}o levava em conta a microestrutura da radia\c{c}\~{a}o e n\~{a}o poderia, portanto, ter validade geral {\rm [...]}.\\
Enquanto Lorentz e Poincar\'{e} consideraram as transforma\c{c}\~{o}es de Lorentz como um postulado separado necess\'{a}rio para derivar a covari\^{a}ncia da teoria eletromagn\'{e}tica, {\rm [...]} Einstein deduziu estas transforma\c{c}\~{o}es a partir de dois axiomas que diziam respeito \`{a} ``natureza do espa\c{c}o e do tempo em geral''.}}
\end{quotation}

O que se viu aqui, portanto, foi que diferentes teorias para o espa\c{c}o e o tempo implicam formas distintas de propaga\c{c}\~{o}es de ondas, como \'{e} o caso para o som e para a luz.

\renewcommand\refname{Refer\^{e}ncias bibliogr\'{a}ficas}

\end{document}